\documentclass{emulateapj-rtx4}
\usepackage{natbib}
\usepackage{graphicx}
\usepackage{makecell}
\shortauthors{Kong et al.}
\begin{document}

\title{Joint Analysis of Energy and RMS Spectra from MAXI J1535-571 with Insight-HXMT}

\author{L. D. Kong$^{1,2}$\textsuperscript{*}, S. Zhang$^{1}$\textsuperscript{*}, Y. P. Chen$^{1}$\textsuperscript{*}, L. Ji$^{3}$, S. N. Zhang$^{1,2}$, Y. R. Yang$^{1}$, L. Tao$^{1}$, X. Ma$^{1}$, J. L. Qu$^{1,2}$, F. J. Lu$^{1}$
% ...
\\
(The $Insight$-HXMT Collaboration)}

\altaffiltext{1}
{Key Laboratory of Particle Astrophysics, Institute of High Energy Physics, Chinese Academy of Sciences, Beijing 100049, China}

\altaffiltext{2}
{University of Chinese Academy of Sciences, Chinese Academy of Sciences, Beijing 100049, China}

\altaffiltext{3}
{Institut f\"ur Astronomie und Astrophysik, Kepler Center for Astro and Particle Physics, Eberhard Karls Universit\"at, 72076 T\"ubingen, Germany}

\altaffiltext{4}
{Department of Astronomy, Beijing Normal University, Beijing 100088, China}

\altaffiltext{5}
{Department of Astronomy, Tsinghua University, Beijing 100084, China}

\altaffiltext{6}
{Department of Physics, Tsinghua University, Beijing 100084, China}

\altaffiltext{7}
{Department of Engineering Physics, Tsinghua University, Beijing 100084, China}

\altaffiltext{8}
{Key Laboratory of Space Astronomy and Technology, National Astronomical Observatories, Chinese Academy of Sciences, Beijing 100012, China}

\altaffiltext{9}
{College of physics Sciences \& Technology, Hebei University, Baoding 071002, Hebei Province, China}

\begin{abstract}
 A new black hole X-ray binary (BHXRB) MAXI J1535-571 was discovered by MAXI during its outburst in 2017. Using observations taken by the first Chinese X-ray satellite, the Hard X-ray Modulation Telescope (dubbed as \emph{Insight}-HXMT), we perform a joint spectral analysis (2-150 keV) in both energy and time domains. The energy spectra provide the essential input for probing the intrinsic Quasi-Periodic Oscillation (QPO) fractional rms spectra (FRS).
 Our results show that during the intermediate state, the energy spectra are in general consistent with those reported by Swift/XRT and NuSTAR. However, the QPO FRS become harder and the FRS residuals may suggest the presence of either an additional power-law component in the energy spectrum or a turn-over in the intrinsic QPO FRS at high energies.
\end{abstract}

\section{Introduction}
A low mass black hole X-ray binary (BHXRB) is composed of a black
hole and a low mass companion star. The black hole can accrete
matter from the companion star via Roche-lobe overflow and form an
optically thick, geometrically thin accretion disk
\citep{Shakura1973}. Because of the hydrogen-ionization thermal
instability in the accretion disk (\citealp{CCL1995};
\citealp{Lasota2001}), the X-ray luminosity increases and an
outburst occurs after a long quiescent period. During an outburst,
the binary system normally transits from a low hard state (LHS), to
the hard/soft intermediate state (HIMS/SIMS), then to the high soft
state (HSS), and finally moves back to LHS. These spectral states
are classified based on the spectral and timing properties
(\citealp{Belloni 2010}; \citealp{RM2006}; \citealp{Motta et al. 2009}).

The energy spectrum of an outburst mainly consists of two component: thermal emission from the standard thin accretion disk and non-thermal emission from the corona/jet. An additional component of the disk reflection is occasionally visible as well, which shows up in the energy spectrum with a  relativistically broadened fluorescent Iron-K$\alpha$ line and a Compton hump above 10 keV (\citealp{Fabian et al. 2000}; \citealp{Miller2007}).
In the LHS, the non-thermal spectrum can also be described by a phenomenological cutoff
power-law, with a spectral index $\Gamma\sim1.6$, and an exponential
cutoff $E_{\rm cut}$ at $50\sim100$ keV \citep{Esin et al. 1997}. A
power density spectrum (PDS) in the LHS normally has a strong
band-limited noise ($30\%\sim40\%$ rms) component accompanied by a
low frequency Quasi-Periodical Oscillation (QPO) occasionally.

As the spectral state moves to the HIMS, the disk emission starts to
dominate and results in a softer spectrum. The band-limited noise
decreases with increasing flux, and the total rms of HIMS
($10\%\sim20\%$) is lower than that in the LHS. Compared to a PDS in
the HIMS where type-C QPOs are normally present, a PDS in the SIMS
has rms lower than $10\%$ and a weak power law component with
presence of a type-A or type-B QPO occasionally. When the spectrum
turns into the HSS, the spectrum can be described by a strong
multi-color thermal blackbody component with a weak power-law tail.
The PDS shows a further weakened power-law noise component.

Such spectral transitions in outbursts of BHXRBs can be explained by
a disk accretion model proposed by \cite{Esin et al. 1997}, who
suggested that the inner disk is truncated and replaced by an
advection-dominated accretion flow in the LHS, and the inner disk
extends to the innermost stable circular orbit (ISCO) in the HSS.
However, the physical origin of QPOs is still in debate. One
possibility is the Lense-Thirring (L-T) precession of the inner
accretion flow due to the frame-dragging effect in the strong
gravitational field (\citealp{Ingram et al. 2009}). Since the QPOs
are highly correlated with the spectral states, the QPO properties
e.g. fractional rms (\citealp{Qu et al. 2010}; \citealp{Rawat et al.
2019}) can provide additional information for diagnosing outburst
evolution jointly with energy spectra.

Since rms is defined as the fractional flux variability, the QPO
fractional rms spectrum (FRS) is supposed to be highly correlated
with the balance between the different energy spectral components.
For sources like MAXI J1535-571 (\citealp{Huang et al. 2018}), XTE
J1859+226 (\citealp{Casella2004}), and H1743-322
(\citealp{Li2013b}), the QPO FRS increases with energy until 20 keV
and remains constant at high energies. The type-C QPOs of these
sources are thought to originate from the corona. However, in GRS
1915+105, the QPO FRS shows a decreasing trend above 20 keV, which
is probably related to a compact jet (\citealp{Rodriguez et al.
2004}). \cite{You2018} performed a numerical simulation based on the
L-T Precession model, and found that the QPO FRS tends to increase
and flatten with energies at larger inclination angles.
\cite{GZ2005} found that the FRS integrated over a broad frequency
range (1/512 to 128 Hz) can be properly recovered by adjusting the
model parameters of the energy spectrum. \cite{SZ2006} reported
anti-correlation between energy spectra derived around QPO
frequencies and an average over time.

MAXI J1535-571 is a new black hole candidate which was discovered by
MAXI/GSC during its outburst in 2017. A series of follow-up
observations were carried out by Swift/BAT, INTEGRAL,
\emph{Insight}-HXMT, NuSTAR and NICER. Low frequency (0.1 $\sim$ 10
Hz) QPOs (type-A, B, C) were detected in spectral states of LHS,
HIMS and SIMS (\citealp{Huang et al. 2018}; \citealp{SK2018};
\citealp{Mereminskiy et al. 2018}; \citealp{Stevens et al. 2018}).
The previous Swift/XRT and NuSTAR observations revealed a system
with an inclination angle around $57^{\circ}$, a black hole spin $a=cJ/GM^{2}\ >
0.84$ and an inner radius $R_{\rm in}\ <\ 2.01R_{\rm ISCO}$ (\citealp{Tao et al.
2018}; \citealp{Xu et al. 2018}). Further researches by \cite{Miller2018} show a near-maximal spin parameter of $a=0.994(2)$ and a disk that extends close to the innermost stable circular orbit, $r/r_{ISCO}=1.08(8)$. Radio observations from the
Australia Telescope Compact Array (ATCA) and MeerKAT showed an
evolution of radio jet during the outburst. (\citealp{Russell2019};
\citealp{Parikh2019}). From the motion of the apparently superluminal knot, \citealp{Russell2019} constrained the jet inclination (at the time of ejection) and speed to $\le$ $41^{\circ}$ and $\ge$ $0.73$ c, respectively. Using the Australian Square Kilometre Array
Pathfinder (ASKAP), \cite{Chauhan2019} constrained a distance of
$4.0_{-0.2}^{+0.2}\ $ kpc to the source by studying the $\rm
H\uppercase\expandafter{\romannumeral1}$ absorption from gas clouds
along the line-of-sight. Here we report the thorough
\emph{Insight}-HXMT observations in a broad energy band from the
beginning of the outburst to the transition towards the soft state
(Section 3.1). The results of our joint analysis of the QPO FRS and
the energy spectra are shown in Section 3.2 \& 3.3. Discussions and
summary are presented in Sections 4 and 5.

\section{Observations and Data analysis}
Following the discovery of MAXI J1535-571 by MAXI/GSC and Swift/BAT,
a series of \emph{Insight}-HXMT Target of Opportunity (ToO)
observations were triggered, which cover a time period of September
6 - 23 in 2017. There is a gap between September 7 and 12 due to an
X9.3 solar flare. Details of these observations are shown in
\cite{Huang et al. 2018}.  We use the same observations for our
joint energy spectra and QPO FRS analysis. The type-C QPOs with rms
larger than 10$\%$ are taken for FRS fitting.

The Hard X-ray Modulation Telescope, also dubbed as \emph{Insight}-HXMT (\citealp{Zhang et al. 2014}), has a broad energy band (1-250 keV) and a large effective area above 20 keV. \emph{Insight}-HXMT consists of three collimated telescopes: the High Energy X-ray Telescope (HE, 18 cylindrical NaI(Tl)/CsI(Na) phoswich detectors), the Medium Energy X-ray Telescope (ME, 1728 Si-PIN detectors), and the Low Energy X-ray Telescope (LE, Swept Charge Device (SCD)), with collecting-area/energy-range of $\sim$5000 $\rm cm^2$/20-250 keV, $\sim$900 $\rm cm^2$/5-30 keV and $\sim$400 $\rm cm^2$/1-10 keV, and typical Field of View (FoV) of $1.6^{\circ}\times6^{\circ}$, $1^{\circ}\times4^{\circ}$ and $1.1^{\circ}\times5.7^{\circ}$ for LE, ME and HE, respectively (\citealp{ZhangSN2019}).

We use the \emph{Insight}-HXMT Data Analysis Software (HXMTDAS)
v2.01 to analyze all data. The data are filtered using the
good-time-interval (GTI) recommended by the \emph{Insight}-HXMT
team; the elevation angle (ELV) is larger than $10^{\circ}$; the
geometric cutoff rigidity (COR) is larger than $8^{\circ}$; the
offset for the point position is smaller than $0.04^{\circ}$; data
are used at least 300 s before and after the South Atlantic Anomaly
(SAA) passage. The energy bands, adopted for energy spectral
analysis are 1-10 keV (LE), 10-27 keV (ME) and 27-150 keV (HE). The
backgrounds are estimated with the official tools: LEBKGMAP ,
MEBKGMAP and HEBKGMAP in version 2.0.6. The XSPEC v12.10.0c software
package (\citealp{Arnaud1996}) is used to perform the spectral
fitting. Uncertainty estimated for each spectral parameter is $90\%$
and a systematic error of 2$\%$ is added.

For timing analysis, we use $powspec$ to produce the PDS from 256 s
intervals with a time resolution of 0.004 s for each observation by
taking Miyamoto normalization (\citealp{Miyamoto et al. 1991}) and
subtract the Poisson noise. The PDS is fitted by XSPEC with multiple
Lorentzians (\citealp{Nowak2000}) and the QPO FRS is estimated by
using $cpflux$ (model in XSPEC) in a frequency range from 1/256 to
125 Hz, for the QPO component with $Q\equiv\nu_{0}/\sigma > 2 $,
where $Q$ is the quality factor, $\nu_{0}$ the centroid frequency of
QPO and $\sigma$ the full width at half maximum (FWHM) of the
Lorentzian function. To produce the QPO FRS, we calculate the QPO
rms in different energy ranges.

\section{Results}
\subsection{Broad-band energy spectrum}
We fit the spectrum with a model composed of a cutoff power-law (non-thermal emission from corona) and a disk blackbody (thermal emission from accretion disk). The absorption due to the Inter Stellar Medium (ISM) $tbabs$ \citep{WAM2000} is also added.
A possible Fe $K_{\rm \alpha}$ line can be seen in the LHS spectrum,
as shown in Figure 1 for ObsID 107. However, we note that the
residual around 6.4 keV might also be due to imprecise background
modeling and calibration. After considering a systematic error of
$2\%$, the reduced $\chi^{2}$ of 1.12 suggests an acceptable fit for
the spectrum. We find that the Fe $K_{\rm \alpha}$ line is hardly
visible when the source enters the HIMS, and the $E_{\rm cut}$
cannot be constrained once the source moves to SIMS.
All parameters are listed in Table 1. Most of the fits have a reduced $\chi^{2}$ around 1, except for ObsIDs 145 and 501 (with the reduced $\chi^{2}$ about $1.4\sim1.5$). As shown in Figure 1, the residuals are mainly at energies around $7\sim10$ and $20\sim30$ keV, where \emph{Insight}-HXMT has relatively larger uncertainties in calibration of LE and ME. We would like to note also that, although the low disk temperature and the $N_{\rm H}$ may have some degeneracy in the spectral fitting, the disk temperatures as listed in Table 1 are in general consistent with those reported by Swift/XRT (\citealp{Tao et al. 2018}).

\begin{figure*}[htbp]
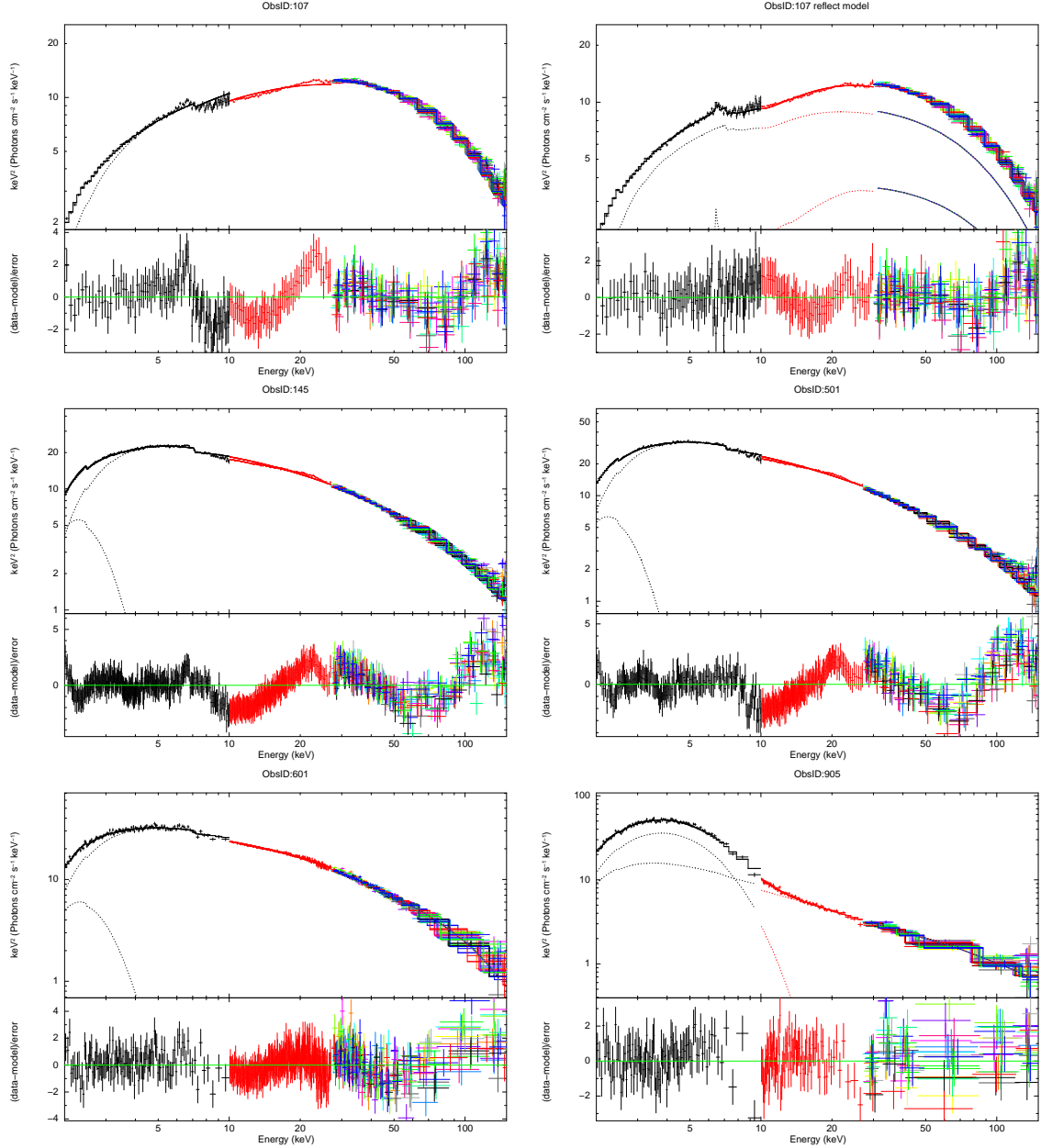

\centering\includegraphics[angle=-90, scale=0.3]{107_fit.eps}
\centering\includegraphics[angle=-90, scale=0.3]{ref_107_model1.eps}
\centering\includegraphics[angle=-90, scale=0.3]{145_fit.eps}
\centering\includegraphics[angle=-90, scale=0.3]{501_fit.eps}
\centering\includegraphics[angle=-90, scale=0.3]{601_fit.eps}
\centering\includegraphics[angle=-90, scale=0.3]{905_fit.eps}
\caption{Spectral fittings with \emph{Insight}-HXMT observations from LE, ME and HE. The black points: LE (2 - 10 keV); the red points: ME (10 - 27 keV); the others are the HE (17 detectors without blind detector No.16, 27 - 150keV). All but the upper right figure: model $tbabs \times (diskbb+cutoffpl)$ for fitting the data from the ObsIDs: 107, 145, 501, 601 and 905. The upper right figure: alternative fitting with reflection model $tbabs \times (diskbb+relxilllpCp+xillverCp)$. The iron line and reflection feature are not significant during HIMS and SIMS.}
\end{figure*}

The evolutions of the spectral parameters are given in Figure 2.
One sees that the thermal component evolves in a trend consistent with that reported previously by \cite{Tao et al. 2018} based on a series of Swift/XRT snapshots at soft X-rays.
However, we notice that $N_{\rm disk}$ (1141.63 on average) is larger than that derived with
Swift/XRT (365.0 on average but with very large uncertainty) during HIMS
(\citealp{Tao et al. 2018}). Also along with the outburst evolving
towards SIMS, the spectrum of the non-thermal component tends to
have larger spectral index and $E_{\rm cut}$ values.

\begin{table*}[ptbptbptb]
\begin{center}
\label{table}
\caption{The spectral fit parameters of MAXI J1535-571 (2$\%$ system err).}
\begin{tabular}{c|c|cccccc|c}
\hline
\hline
$^{c}$ObsID & Time (s) & $N_{\rm H}$ &  $kT_{\rm in}$ & $^{b}D_{\rm norm}$ & $\Gamma$ & $E_{\rm cut} $ & $^{b}C_{\rm norm}$ & $\chi_{\rm red}^{2}(d.o.f)$
\\
& & $(10^{22}\rm cm^{-2})$ & (keV) & $(10^{5})$ & & (keV) & &
\\
\hline
106 & 2017-09-06T23:33:06 & $ 3.3_{-0.3}^{+0.4} $ & $ 0.42_{-0.02}^{+0.02} $ & $ 0.3_{-0.1}^{+0.2} $ & $ 1.46_{-0.01}^{+0.01} $ & $ 48.0_{-0.5}^{+0.6} $ & $ 3.8_{-0.1}^{+0.1} $ & 1.09(2819)
\\
107 & 2017-09-07T02:43:10 & $ 3.5_{-0.1}^{+0.3} $ & $ 0.42_{-0.01}^{+0.01} $ & $ 0.31_{-0.06}^{+0.1} $ & $ 1.47_{-0.01}^{+0.01} $ & $ 47.9_{-0.5}^{+0.6} $ & $ 3.98_{-0.07}^{+0.1} $ & 1.12(2819)
\\
108 & 2017-09-07T05:54:06 & $ 2.4_{-0.4}^{+0.3} $ & $ 0.4_{-0.1}^{+0.03} $ & $ 0.08_{-0.06}^{+0.09} $ & $ 1.40_{-0.01}^{+0.02} $ & $ 44.1_{-0.9}^{+0.7} $ & $ 3.5_{-0.1}^{+0.1} $ & 1.08(2819)
\\
144 & 2017-09-12T10:38:15 & $ 4.2_{-0.3}^{+0.2} $ & $ 0.34_{-0.01}^{+0.01} $ & $ 7.9_{-2.1}^{+1.4} $ & $ 2.26_{-0.01}^{+0.01} $ & $ 63.1_{-1.3}^{+1.5} $ & $ 42.3_{-1.1}^{+1.1} $ & 1.14(2819)
\\
145 & 2017-09-12T13:58:12 & $ 5.1_{-0.1}^{+0.2} $ & $ 0.32_{-0.01}^{+0.01} $ & $ 26.2_{-3.1}^{+4.1} $ & $ 2.37_{-0.01}^{+0.01} $ & $ 76.3_{-0.7}^{+1} $ & $ 52.2_{-0.7}^{+0.9} $ & 1.50(2819)
\\
301 & 2017-09-15T04:48:00 & $ 5.1_{-0.3}^{+0.2} $ & $ 0.32_{-0.01}^{+0.01} $ & $ 27.7_{-4.4}^{+5.8} $ & $ 2.21_{-0.01}^{+0.01} $ & $ 57.1_{-0.9}^{+0.7} $ & $ 46.6_{-1}^{+0.8} $ & 1.17(2811)
\\
401 & 2017-09-16T06:15:29 & $ 5.0_{-0.2}^{+0.2} $ & $ 0.32_{-0.01}^{+0.01} $ & $ 29.4_{-4.1}^{+4.8} $ & $ 2.38_{-0.004}^{+0.01} $ & $ 68.6_{-0.2}^{+0.9} $ & $67.08_{-1}^{+1.7} $ & 1.24(2819)
\\
501 & 2017-09-17T06:07:38 & $ 4.9_{-0.2}^{+0.1} $ & $ 0.31_{-0.01}^{+0.01} $ & $ 34.8_{-4.5}^{+4.4} $ & $ 2.50_{-0.01}^{+0.01} $ & $ 79.5_{-1.6}^{+1.5} $ & $ 91.4_{-2.6}^{+1.5} $ & 1.42(2819)
\\
601 & 2017-09-18T02:48:54 & $ 4.4_{-0.4}^{+0.4} $ & $ 0.36_{-0.02}^{+0.01} $ & $ 8.1_{-2.4}^{+3.7} $ & $ 2.37_{-0.01}^{+0.01} $ & $ 60.6_{-0.8}^{+1.4} $ & $ 74.3_{-2.4}^{+2.7} $ & 1.07(2664)
\\
901 & 2017-09-21T02:26:26 & $ 3.11_{-0.04}^{+0.03} $ & $ 1.27_{-0.01}^{+0.01} $ & $ ^{\mathrm{a}}1.90_{-0.07}^{+0.09} $ & $ 2.60_{-0.01}^{+0.01} $ & $ 94.5_{-3.3}^{+4.8} $ & $ 83.6_{-4.1}^{+2.6} $ & 1.03(2819)
\\
902 & 2017-09-21T06:00:41 & $ 3.44_{-0.05}^{+0.04} $ & $ 1.26_{-0.01}^{+0.01} $ & $ ^{\mathrm{a}}1.77_{-0.07}^{+0.03} $ & $ 2.76_{-0.01}^{+0.02} $ & $ <147 $ & $ 117.7_{-4.1}^{+4.9} $ & 1.18(2819)
\\
903 & 2017-09-21T09:21:07 & $ 3.03_{-0.03}^{+0.07} $ & $ 1.29_{-0.01}^{+0.01} $ & $ ^{\mathrm{a}}1.62_{-0.05}^{+0.06} $ & $ 2.55_{-0.02}^{+0.02} $ & $ 82.9_{-5.3}^{+4.2} $ & $ 80.0_{-2.5}^{+4.9} $ & 0.91(2736)
\\
905 & 2017-09-21T21:40:13 & $ 2.97_{-0.07}^{+0.1} $ & $ 1.19_{-0.01}^{+0.004} $ & $ ^{\mathrm{a}}3.7_{-0.2}^{+0.2} $ & $ 2.82_{-0.01}^{+0.03} $ & $ - $ & $ 60.3_{-2.7}^{+5.7} $ & 1.09(2539)
\\
906 & 2017-09-22T00:46:04 & $ 2.85_{-0.05}^{+0.08} $ & $ 1.17_{-0.01}^{+0.01} $ & $ ^{\mathrm{a}}4.4_{-0.1}^{+0.2} $ & $ 2.68_{-0.04}^{+0.03} $ & $ - $ & $ 39.2_{-3.3}^{+3.5} $ & 1.03(2771)
\\
907 & 2017-09-22T04:09:23 & $ 3.13_{-0.04}^{+0.06} $ & $ 1.16_{-0.01}^{+0.01} $ & $ ^{\mathrm{a}}4.1_{-0.2}^{+0.1} $ & $ 2.81_{-0.01}^{+0.01} $ & $ - $ & $ 64.9_{-2.1}^{+3.9} $ & 1.09(2645)
\\
908 & 2017-09-22T07:32:40 & $ 2.98_{-0.05}^{+0.06} $ & $ 1.20_{-0.01}^{+0.01} $ & $ ^{\mathrm{a}}3.49_{-0.09}^{+0.08} $ & $ 2.71_{-0.01}^{+0.02} $ & $ - $ & $ 59.9_{-2.2}^{+3.1} $ & 1.01(2810)
\\
911 & 2017-09-22T22:59:53 & $ 2.64_{-0.05}^{+0.09} $ & $ 1.18_{-0.01}^{+0.01} $ & $ ^{\mathrm{a}}4.5_{-0.1}^{+0.2} $ & $ 2.40_{-0.02}^{+0.06} $ & $ 51.8_{-3.8}^{+7.3} $ & $ 20.6_{-1.7}^{+3.3} $ & 1.06(2300)
\\
912 & 2017-09-23T02:15:02 & $ 2.79_{-0.04}^{+0.06} $ & $ 1.18_{-0.01}^{+0.01} $ & $ ^{\mathrm{a}}4.2_{-0.1}^{+0.2} $ & $ 2.61_{-0.03}^{+0.03} $ & $ - $ & $ 37.1_{-2.5}^{+3} $ & 1.10(2815)
\\
913 & 2017-09-23T05:43:54 & $ 2.89_{-0.04}^{+0.06} $ & $ 1.19_{-0.01}^{+0.004} $ & $ ^{\mathrm{a}}3.80_{-0.07}^{+0.2} $ & $ 2.70_{-0.02}^{+0.02} $ & $ - $ & $ 50.0_{-2.7}^{+2} $ & 1.05(2788)
\\
914 & 2017-09-23T09:03:56 & $ 2.97_{-0.05}^{+0.06} $ & $ 1.16_{-0.01}^{+0.004} $ & $ ^{\mathrm{a}}4.38_{-0.09}^{+0.1} $ & $ 2.77_{-0.02}^{+0.03} $ & $ - $ & $ 51.5_{-3.1}^{+4.3} $ & 1.05(2738)
\\
915 & 2017-09-23T12:25:30 & $ 2.87_{-0.05}^{+0.03} $ & $ 1.19_{-0.01}^{+0.01} $ & $ ^{\mathrm{a}}3.66_{-0.08}^{+0.09} $ & $ 2.65_{-0.02}^{+0.02} $ & $ - $ & $ 48.4_{-3.1}^{+2} $ & 1.13(2827)
\\
916 & 2017-09-23T15:46:21 & $ 3.26_{-0.07}^{+0.07} $ & $ 1.22_{-0.01}^{+0.01} $ & $ ^{\mathrm{a}}2.89_{-0.04}^{+0.04} $ & $ 2.85_{-0.02}^{+0.02} $ & $ - $ & $ 90.8_{-5.1}^{+6.5} $ & 1.12(2819)
\\
917 & 2017-09-23T18:56:22 & $ 3.50_{-0.08}^{+0.02} $ & $ 1.25_{-0.01}^{+0.01} $ & $ ^{\mathrm{a}}2.3_{-0.1}^{+0.1} $ & $ 2.95_{-0.02}^{+0.01} $ & $ - $ & $ 121.2_{-7.5}^{+2.5} $ & 1.10(2769)
\\
918 & 2017-09-23T22:06:36 & $ 2.44_{-0.05}^{+0.08} $ & $ 1.22_{-0.01}^{+0.01} $ & $ ^{\mathrm{a}}3.68_{-0.08}^{+0.2} $ & $ 2.22_{-0.06}^{+0.05} $ & $ 37.7_{-4.7}^{+4.1} $ & $ 16.7_{-1.9}^{+2.4} $ & 0.97(1928)
\\
\hline
\end{tabular}
\end{center}
\begin{list}{}{}
\item[${\mathrm{a}}$]{: $(10^{3})$}
\item[${\mathrm{b}}$]{: The normalization of $diskbb$ model.}
\item[${\mathrm{c}}$]{: P011453500XXX}
\end{list}
\end{table*}
\begin{figure*}[htbp]
\centering\includegraphics[scale=1]{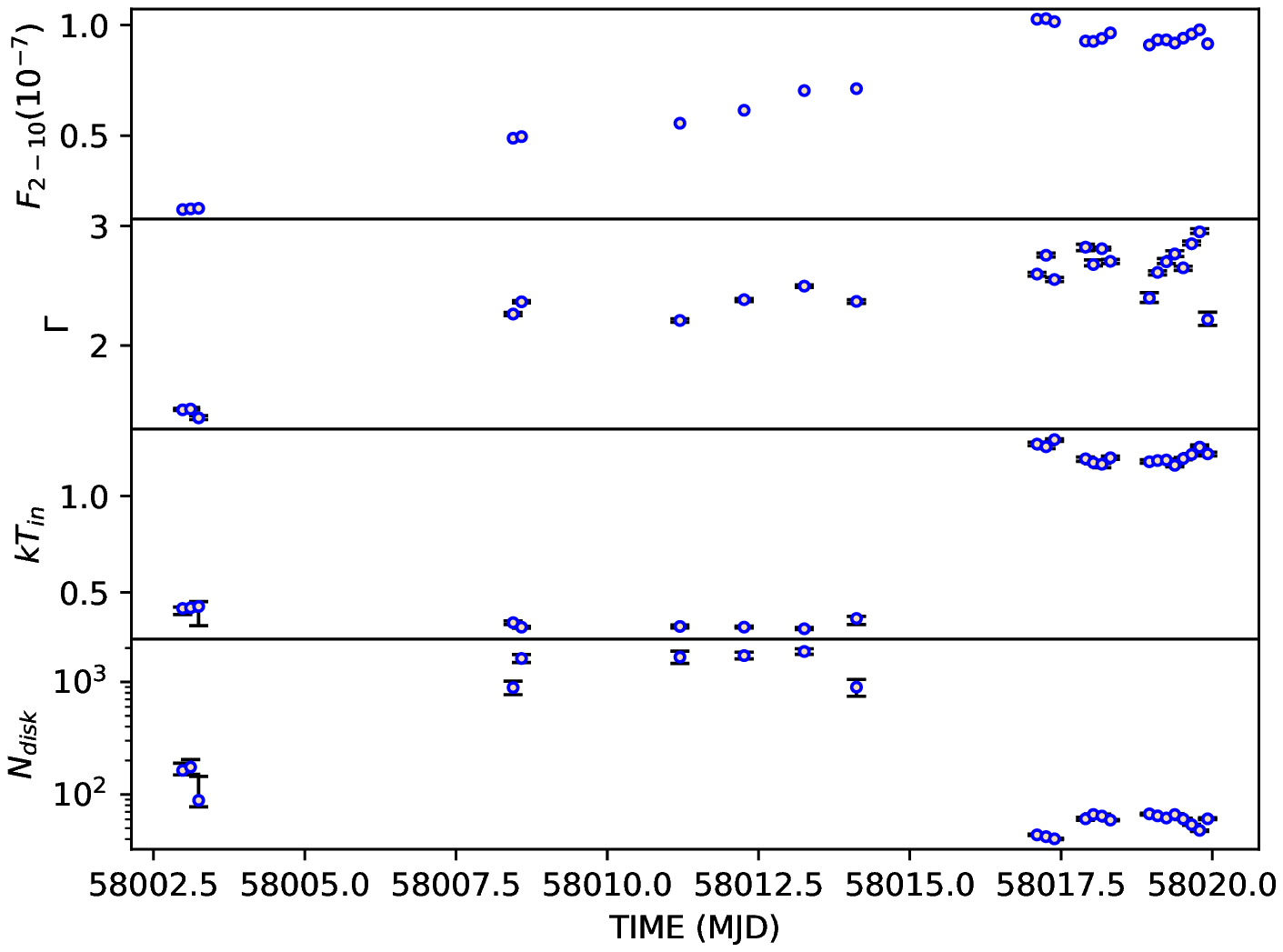}
\caption{The evolution of \emph{Insight}-HXMT spectral parameters. The parameters
are taken from Table 1. $F_{2-10}$ is the flux in 2-10 keV in unit
of $10^{-7} \rm ergs\ \rm cm^{-2}\ s^{-1}$; $\Gamma$ is the photon
index; $T_{in}$ is the inner disk temperature in unit of keV;
$N_{\rm disk} \equiv R_{\rm in} \times D_{10}^{-1} \times \sqrt{\rm
cos\theta}$ is the $\sqrt{D_{\rm norm}}$.}
\end{figure*}

In Figure 1, we show the spectra of the observations taken in the LHS, HIMS and SIMS. \emph{Insight}-HXMT ObsID 107 is about 16 hours earlier than the corresponding NuSTAR observation (2017-09-07 18:41:09 UTC). A prominent reflection component was detected in the NuSTAR observation \citep{Xu et al. 2018}. We use model $tbabs \times (diskbb+relxilllpCp+xillverCp)$ (denoted as M1) to fit the energy spectrum observed by \emph{Insight}-HXMT in ObsID 107, in which relxilllpCp is a lamp-post model in the relxill model family \citep{Dauser2014}. 
The parameter $fixReflFrac$ is a switch, which is set to 1 to use the predicted value of the current parameter configuration in the lamp post geometry. Alternatively, we use model $tbabs \times (diskbb+nthcomp)$ (denoted as M2) to fit ObsID 107 for comparison.

\begin{table}[ptbptbptb]
\begin{center}
\label{table}
\caption{Fit results of ObsID: 107}
\begin{tabular}{cccc}
\hline
\hline
Component & Parameters & M1 & M2
\\
\hline
Tbabs & $N_{\rm H}\ (10^{22}\ \rm cm^{-2})$ & $4.67_{-0.09}^{+0.05}$ & $7.0_{-0.3}^{+0.2}$
\\
\\
Diskbb & $kT_{\rm in}\ (\rm keV)$ & $0.29_{-0.004}^{+0.01}$ & $0.32_{-0.01}^{+0.002}$
\\
& Norm $(10^{5})$ & $8.1_{-0.7}^{+0.7}$ & $15.6_{-2.1}^{+5.6}$
\\
\\
nthComp & $\Gamma$ & - & $1.88_{-0.006}^{+0.02}$
\\
& $kT_{\rm e}\ (\rm keV)$ & - & $25.72_{-0.29}^{+0.16}$
\\
& Norm & - & $4.8_{-0.1}^{+0.2}$
\\
\\
relxilllpCph & $h\ (r_{\rm g})$ & $6.3_{-0.4}^{+0.3}$ & -
\\
& $a\ (\rm cJ/GM^{2})$ & $0.7_{-0.3}^{+0.2}$ & -
\\
& $R_{\rm in}\ (r_{\rm ISCO})$ & $1.20_{-0.01}^{+0.03}$ & -
\\
& $i\ (^{\circ})$ & $53.9_{-2.3}^{+1.8}$ & -
\\
& $\Gamma$ & $1.88_{-0.01}^{+0.001}$ & -
\\
& $\rm log\xi\ (\rm log\ [\rm erg\ \rm cm\ \rm s^{-1}])$ & $3.38_{-0.02}^{+0.04}$ & -
\\
& $A_{\rm Fe}\ (\rm solar)$ & $0.50_{-0.001}^{+0.06}$ & -
\\
& $kT_{\rm e}\ (\rm keV)$ & $48.6_{-1.9}^{+0.5}$ & -
\\
& Norm & $0.09_{-0.01}^{+0.02}$ & -
\\
\\
xillverCp & $\rm log\xi\ (\rm log\ [\rm erg\ \rm cm\ s-1])$ & $2.29_{-0.001}^{+0.02}$ & -
\\
& Norm & $0.018_{-0.002}^{+0.002}$ & -
\\
\hline
& $\chi_{\rm red}^{2}/d.o.f$ & 0.96/2822 & 1.64/2830
\\
\hline
\end{tabular}
\end{center}
\begin{list}{}{}
\item[Note]{: Uncertainties are reported at the 90\% confidence interval and were computed using MCMC (Markov Chain Monte Carlo) of length 100,000.}
\end{list}
\end{table}
The results (see Table 2) show consistence between the measurements from the two telescopes except for the temperature of the disk, which may due to a broader energy coverage of \emph{Insight}-HXMT and the degeneracy in spectral fitting at low energies between disk temperature and $N_{\rm H}$. Apart from the NuSTAR observation, there exist several NuSTAR snapshots very close in time to the \emph{Insight}-HXMT observations (ObsID 145 and 501). We perform a similar analysis and find that the $E_{\rm cut}$ derived from the two telescopes is consistent: $63.13_{-1.30}^{+1.46}$ and $79.53_{-1.62}^{+1.52}$ from \emph{Insight}-HXMT OnsID 145 and 501 vs $62.88_{-2.58}^{+2.80}$ and $57.70_{-3.58}^{+4.04}$ from the two NuSTAR observations. Hence we speculate that there may be a change of $E_{\rm cut}$ within a few hours on September 17 when the source was entering the HIMS.

\subsection{Type-C QPOs}
Following \cite{Huang et al. 2018}, for the \emph{Insight}-HXMT observations of MAXI J1535-571, we take nine ObsIDs in which Type-C QPOs are detected. The PDS was fitted with several Lorentzians. The background contribution to the QPO FRS is considered in the definition of  rms (\citealp{Bu et al. 2015}):
\begin{equation}
\rm rms = \sqrt{\emph{P}}\times(\emph{S+B})/\emph{S},
\end{equation}
where $S$ and $B$ represent the mean count rates of source and background respectively, and $P$ is the power calculated with integration of the QPO lorentzian function over the frequency range $1/256\sim125$ Hz. The properties of the Type-C QPOs as shown in Table 3 are consistent with those reported previously by \cite{Huang et al. 2018} by using the same \emph{Insight}-HXMT observations.
\begin{table}[ptbptbptb]
\begin{center}
\label{table}
\caption{Low-Frequency QPO properties}
\begin{tabular}{c|c|ccc}
\hline
\hline
ObsID & Type & QPO $\nu$ & $\sigma$ & rms
\\
& & (Hz) & & (\%)
\\
\hline
P011453500144 & C & $2.53_{-0.02}^{+0.02}$ & $0.25_{-0.04}^{+0.04}$ & $ 9.2_{-2.9}^{+2.9}$
\\
P011453500145 & C & $2.61_{-0.01}^{+0.01}$ & $0.41_{-0.03}^{+0.03}$ & $10.1_{-2.2}^{+2.2}$
\\
P011453500301 & C & $2.02_{-0.01}^{+0.01}$ & $0.20_{-0.03}^{+0.03}$ & $10.0_{-3.2}^{+2.3}$
\\
P011453500401 & C & $2.76_{-0.01}^{+0.01}$ & $0.22_{-0.02}^{+0.02}$ & $10.5_{-2.1}^{+2.8}$
\\
P011453500501 & C & $3.33_{-0.01}^{+0.01}$ & $0.31_{-0.03}^{+0.03}$ & $11.2_{-3.2}^{+2.7}$
\\
P011453500601 & C & $3.32_{-0.03}^{+0.03}$ & $0.29_{-0.05}^{+0.05}$ & $10.7_{-3.8}^{+3.7}$
\\
P011453500901 & C & $9.21_{-0.04}^{+0.04}$ & $0.60_{-0.06}^{+0.06}$ & $10.7_{-3.2}^{+2.8}$
\\
P011453500902 & C & $9.35_{-0.03}^{+0.04}$ & $0.84_{-0.09}^{+0.06}$ & $11.6_{-3.8}^{+3.3}$
\\
P011453500903 & C & $8.87_{-0.45}^{+0.09}$ & $0.09_{-0.18}^{+0.15}$ & $11.0_{-3.4}^{+5.6}$
\\
\hline
\end{tabular}
\end{center}
\begin{list}{}{}
\item[Note]{: The properties of the QPOs was get from the PDS in a narrow energy range (27.4 $\sim$ 31.2 keV) for comparing with the results in \cite{Huang et al. 2018} (Tabel 2, which get from a broad energy range (6 $\sim$ 38 keV))}
\end{list}
\end{table}
During the spectral evolution from HIMS to SIMS, the centroid frequency of QPO increases in general except for two significant drops in ObsID 301 and 903.
\subsection{Joint fitting of energy spectrum and QPO FRS}
The QPO FRS is obtained from integrating over the lorentzian function which represents the QPO component. The result of QPO FRS is shown in Figure 6a, where the type-C QPOs in HIMS (ObsIDs 144-601) and SIMS (ObsIDs 901-903) have an overall similar evolution trend.
The rms is rather small at lower energies and increases gradually with energy until reaching a flat top at energy above roughly 10 keV.
Since the rms is defined as the fractional variability of the flux, the detected QPO FRS as shown in Figure 6a is the composition of rms contributions from different energy spectral components. For example, the FRS at soft X-rays and hard X-rays may be dominated by contributions from disk and corona respectively.
To distinguish this, we need to perform the joint fitting of QPO FRS and energy spectrum, where the different spectral components of the latter can provide the necessary input for investigating the composition of the former.

The model used to fit the FRS is
\begin{equation}
\rm rms\equiv \sigma(\emph E) \times \emph F_{\rm c}/\emph F_{\rm t},
\end{equation}
where $F_{\rm t}$ is the time-averaged total flux, $\sigma(E)$ is a function for the intrinsic FRS, and $F_{\rm c}$ represents the flux that gives contribution to QPO rms.
In the joint fitting with FRS, the energy spectral components, diskbb and cutoff power-law, were taken for fitting the thermal and non-thermal emissions, respectively.
We hence take
\begin{equation}
F_{\rm t} = F_{\rm d}+F_{\rm pl},
\end{equation}
\begin{equation}
F_{\rm c} = F_{\rm pl},
\end{equation}
where $F_{\rm d}$ and $F_{\rm pl}$ represent the flux of the diskbb component and cutoff power-law component in the energy spectra, respectively.
The joint fitting in Xspec is to take the energy spectral inputs from three detectors of \emph{Insight}-HXMT (LE, ME and HE), and the QPO FRS. A model to cover all these inputs is structured as, model (which set in Xspec) = $\rm Constant1 \times tbabs \times (diskbb+cutoffpl)+Constant2 \times rms\ model$ (i.e., Eq.2). For LE, ME and HE energy spectra, the Constant1=1 and Constant2=0, but Constant1=0 and Constant2=1 for QPO FRS.

The first trial (model-1), which means only non-thermal contribution to FRS, is considered as
\begin{equation}
\rm rms = \sigma(\emph E) \times \frac{\emph F_{\rm pl}(\Gamma,\emph E_{\rm cut},\emph{n}_{1})}{\emph F_{\rm d}(\emph T,\emph{n}_{2})+\emph F_{\rm pl}(\Gamma,\emph E_{\rm cut},\emph{n}_{1})},
\end{equation}
where $n_{1}$ is the $C_{\rm norm}$, and $n_{2}$ is the $D_{\rm norm}$ from model \emph{diskbb} and model \emph{cutoffpl}.
\begin{figure}[htbp]
\centering\includegraphics[scale=0.5]{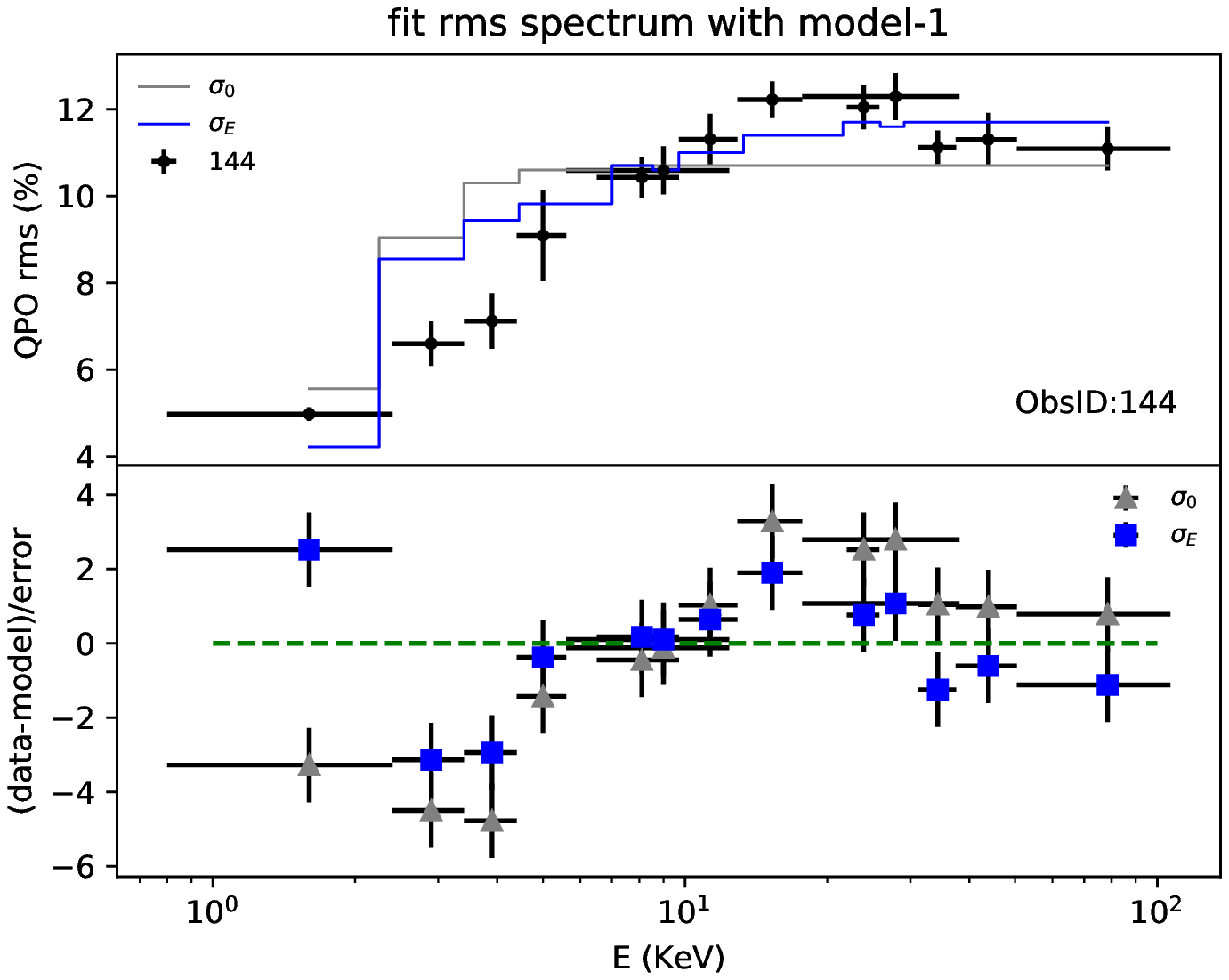}
\caption{The top panel shows the QPO FRS fitting of ObsID: 144. The grey line represents model-1 with $\sigma(E) \equiv \sigma_{0}$ and it is independent of energy. The blue one shows model-1 with an energy function $\sigma(E)$. The bottom panel shows the residuals of fittings.}
\end{figure}
We first try $\sigma(E)$ with an energy independent constant in joint fitting, and find that the spectrum cannot be well fitted, as shown in Figure 3 for one example from ObsID 144, where the grey line and points show large residual in the lower energy end. We hence take an energy-dependent function $\sigma(E)$ instead, which is in the following form:
\begin{equation}
\sigma(E) \equiv \frac{\sigma_{0}}{1+e^{-k(E-E_{0})}}.
\end{equation}
Here $\sigma_{0}$ is the maximum value of $\sigma(E)$, $k$ controls the steepness of this function, and $E_{0}$ is the sigmoid's midpoint. With this function, as shown for the blue line and points in Figure 3, the FRS fitting is improved and the derived parameters are listed in Table 4, and the FRS residuals are shown in Figure 5 with grey triangle points.
\begin{table*}[ptbptbptb]
\begin{center}
\label{table}
\caption{The joint fit without considering disk contribution by model-1:}
\begin{tabular}{c|cccccc|ccc|c}
\hline
\hline
 ObsID & $N_{\rm H}$ &  $kT_{\rm in} $ & $^{b}n_{2}$ & $\Gamma$ & $E_{\rm cut} $ & $^{b}n_{1}$ & $\sigma_{0}$ & k & $E_{0}$ & $\chi_{\rm red}^{2}(d.o.f)$
 \\
 & $(10^{22}\rm cm^{-2})$ & (keV) & $(10^{5})$ & & (keV) & & & & (keV) &
 \\
 \hline
144 & $ 3.49_{-0.27}^{+0.02} $ & $ 0.29_{-0.01}^{+0.01} $ & $ 12.67_{-3.02}^{+2.33} $ & $ 2.25_{-0.02}^{+0.01} $ & $ 61.93_{-1.94}^{+0.23} $ & $ 40.45_{-1.79}^{+0.12} $ & $11.73_{-0.53}^{+0.40}$ & $0.18_{-0.01}^{+0.06}$ & $ -4.23_{-1.60}^{+1.65} $ & 1.17(2829)
\\
145 & $ 5.04_{-0.16}^{+0.17} $ & $ 0.32_{-0.01}^{+0.01} $ & $ 25.87_{-3.41}^{+3.61} $ & $ 2.36_{-0.01}^{+0.01} $ & $ 76.06_{-1.25}^{+0.98} $ & $ 51.99_{-1.13}^{+0.63} $ & $12.58_{-0.23}^{+0.41}$ & $0.25_{-0.07}^{+0.08}$ & $ -2.34_{-2.36}^{+2.86} $ & 1.51(2829)
\\
301 & $ 5.10_{-0.60}^{+0.12} $ & $ 0.32_{-0.01}^{+0.01} $ & $ 29.83_{-9.07}^{+1.92} $ & $ 2.21_{-0.02}^{+0.01} $ & $ 57.24_{-1.17}^{+0.09} $ & $ 46.81_{-2.06}^{+0.08} $ & $11.79_{-0.14}^{+0.67}$ & $0.14_{-0.02}^{+0.05}$ & $ -0.28_{-1.40}^{+2.13} $ & 1.20(2821)
\\
401 & $ 3.91_{-0.06}^{+0.15} $ & $ 0.34_{-0.01}^{+0.02} $ & $ 7.55_{-2.55}^{+1.06} $ & $ 2.31_{-0.01}^{+0.01} $ & $ 62.21_{-0.28}^{+1.14} $ & $ 57.11_{-0.22}^{+1.47} $ & $12.81_{-0.39}^{+0.23}$ & $0.15_{-0.03}^{+0.05}$ & $ -1.22_{-2.36}^{+0.51} $ & 1.32(2829)
\\
501 & $ 4.84_{-0.17}^{+0.14} $ & $ 0.31_{-0.01}^{+0.01} $ & $ 35.48_{-6.45}^{+3.94} $ & $ 2.50_{-0.01}^{+0.01} $ & $ 79.40_{-1.17}^{+1.11} $ & $ 91.12_{-1.72}^{+1.73} $ & $13.71_{-0.19}^{+0.35}$ & $0.30_{-0.06}^{+0.06}$ & $2.28_{-0.72}^{+0.64}$ & 1.43(2829)
\\
601 & $ 4.44_{-0.36}^{+0.07} $ & $ 0.36_{-0.02}^{+0.01} $ & $ 8.48_{-3.69}^{+2.70} $ & $ 2.37_{-0.01}^{+0.01} $ & $ 60.62_{-1.74}^{+1.00} $ & $ 74.38_{-2.18}^{+1.30} $ & $13.27_{-0.35}^{+0.30}$ & $0.34_{-0.09}^{+0.28}$ & $2.12_{-1.34}^{+1.12}$ & 1.07(2672)
\\
901 & $ 3.11_{-0.04}^{+0.02} $ & $ 1.27_{-0.01}^{+0.01} $ & $ ^{\mathrm{a}}1.90_{-0.08}^{+0.03} $ & $ 2.60_{-0.01}^{+0.01} $ & $ 94.62_{-2.54}^{+2.52} $ & $ 83.72_{-1.60}^{+1.58} $ & $14.65_{-0.40}^{+0.46}$ & $0.48_{-0.10}^{+0.13}$ & $3.76_{-0.45}^{+0.21}$ & 1.04(2828)
\\
902 & $ 3.44_{-0.03}^{+0.02} $ & $ 1.26_{-0.01}^{+0.01} $ & $ ^{\mathrm{a}}1.77_{-0.08}^{+0.06} $ & $ 2.76_{-0.01}^{+0.01} $ & $ <147 $ & $ 117.70_{-2.00}^{+2.59} $ & $15.15_{-0.33}^{+0.29}$ & $0.42_{-0.06}^{+0.08}$ & $5.85_{-0.33}^{+0.31}$ & 1.20(2828)
\\
903 & $ 3.04_{-0.05}^{+0.05} $ & $ 1.29_{-0.01}^{+0.01} $ & $ ^{\mathrm{a}}1.63_{-0.05}^{+0.08} $ & $ 2.55_{-0.02}^{+0.02} $ & $ 82.19_{-6.22}^{+3.35} $ & $ 79.85_{-3.67}^{+2.97} $ & $14.23_{-0.25}^{+0.66}$ & $1.28_{-0.05}^{+0.50}$ & $3.41_{-0.22}^{+0.11}$ & 0.94(2744)
\\
\hline
\end{tabular}
\end{center}
\begin{list}{}{}
\item[${\mathrm{a}}$]{: $(10^{3})$}
\item[${\mathrm{b}}$]{: The normalization of $diskbb$ model.}
\end{list}
\end{table*}

However, with model-1 the residuals at lower energies still exist and hence we introduce model-2 with another rms contribution from thermal emission:
\begin{equation}
\rm rms = \sigma_{1}(\emph E) \times\ \emph f_{1}+ \sigma_{2}(\emph E) \times\ \emph f_{2},
\end{equation}
\begin{equation}
f_{1} = \frac{\emph F_{\rm pl}(\Gamma,\emph E_{\rm cut},\emph n_{1})}{\emph F_{\rm d}(\emph T,\emph n_{2})+\emph F_{\rm pl}(\Gamma,\emph E_{\rm cut},\emph n_{1})},
\end{equation}
\begin{equation}
f_{2} = \frac{\emph F_{\rm d}(\emph T,\emph n_{2})}{\emph F_{\rm d}(\emph T,\emph n_{2})+\emph F_{\rm pl}(\Gamma,\emph E_{\rm cut},\emph n_{1})},
\end{equation}
\begin{equation}
\sigma_{1}(\emph E) = \frac{\sigma_{0}}{1+e^{-k(E-E_{0})}},
\end{equation}
\begin{equation}
\sigma_{2}(\emph E) = \sigma_{\rm disk}.
\end{equation}
Again, we try $\sigma_{2}(E)$ in model-2 with a constant $\sigma_{\rm disk}$, and $\sigma_{1}(E)$ the same as $\sigma(E)$ in model-1. The results with model-2 are listed in Table 5, and the residuals are shown in Figure 5 with blue triangle points. One can see in Table 5 that the thermal contributions to the FRS are visible in HIMS but not in SIMS.
\begin{table*}[ptbptbptb]
\begin{center}
\label{table}
\caption{The joint fit with disk component by model-2:}
\begin{tabular}{c|cccp{1cm}<{\centering}p{1.2cm}<{\centering}p{1.8cm}<{\centering}|p{1cm}<{\centering}p{1cm}<{\centering}p{1cm}<{\centering}p{1.1cm}<{\centering}|p{1.3cm}<{\centering}}
\hline
\hline
 ObsID & $N_{\rm H}$ &  $kT_{\rm in} $ & $^{b}n_{2}$ & $\Gamma$ & $E_{\rm cut} $ & $^{b}n_{1}$ & $\sigma_{0}$ & $k$ & $E_{0}$ & $\sigma_{\rm disk}$ & $\chi_{\rm red}^{2}(d.o.f)$
 \\
 & $(10^{22}\rm cm^{-2})$ & (keV) & $(10^{5})$ & & (keV) & & & & (keV) & &
 \\
 \hline
144 & $ 4.14_{-0.23}^{+0.20} $ & $ 0.34_{-0.01}^{+0.01} $ & $ 7.66_{-1.37}^{+0.86} $ & $ 2.26_{-0.01}^{+0.01} $ & $ 62.96_{-1.40}^{+0.96} $ & $ 42.15_{-0.89}^{+0.63} $ & $11.67_{-0.34}^{+0.37}$ & $0.36_{-0.09}^{+0.12}$ & $2.14_{-0.84}^{+0.88}$ & $4.77_{-0.86}^{+1.09}$ & 1.14(2828)
\\
145 & $ 5.07_{-0.22}^{+0.03} $ & $ 0.32_{-0.01}^{+0.01} $ & $ 26.18_{-3.70}^{+2.21} $ & $ 2.37_{-0.01}^{+0.02} $ & $ 76.29_{-1.21}^{+0.52} $ & $ 52.24_{-1.17}^{+0.41} $ & $12.54_{-0.22}^{+0.33}$ & $0.43_{-0.15}^{+0.12}$ & $1.43_{-1.38}^{+0.94}$ & $4.90_{-1.42}^{+1.26}$ & 1.50(2828)
\\
301 & $ 5.11_{-0.22}^{+0.11} $ & $ 0.32_{-0.01}^{+0.02} $ & $ 27.61_{-4.22}^{+4.38} $ & $ 2.21_{-0.01}^{+0.01} $ & $ 57.07_{-0.80}^{+1.05} $ & $ 46.61_{-0.70}^{+0.65} $ & $11.68_{-0.52}^{+0.60}$ & $0.25_{-0.08}^{+0.14}$ & $0.94_{-1.37}^{+0.91}$ & $5.10_{-0.86}^{+1.03}$ & 1.17(2820)
\\
401 & $ 4.96_{-0.14}^{+0.15} $ & $ 0.32_{-0.02}^{+0.01} $ & $ 28.69_{-3.03}^{+3.85} $ & $ 2.38_{-0.01}^{+0.01} $ & $ 68.38_{-1.35}^{+1.56} $ & $ 66.82_{-1.37}^{+1.58} $ & $12.65_{-0.32}^{+0.18}$ & $0.30_{-0.04}^{+0.05}$ & $2.97_{-0.38}^{+0.36}$ & $5.40_{-0.54}^{+0.49}$ & 1.25(2828)
\\
501 & $ 4.88_{-0.13}^{+0.17} $ & $ 0.31_{-0.01}^{+0.01} $ & $ 34.81_{-5.37}^{+4.34} $ & $ 2.50_{-0.01}^{+0.01} $ & $ 79.53_{-1.17}^{+1.38} $ & $ 91.37_{-1.04}^{+1.43} $ & $13.68_{-0.20}^{+0.39}$ & $0.38_{-0.05}^{+0.07}$ & $3.29_{-0.72}^{+0.64}$ & $5.54_{-1.98}^{+1.62}$ & 1.43(2828)
\\
601 & $ 4.44_{-0.15}^{+0.05} $ & $ 0.36_{-0.02}^{+0.01} $ & $ 8.48_{-0.22}^{+0.50} $ & $ 2.37_{-0.01}^{+0.03} $ & $ 60.62_{-0.64}^{+4.51} $ & $ 74.38_{-1.84}^{+0.80} $ & $13.26_{-1.02}^{+0.19}$ & $0.34_{-0.01}^{+0.03}$ & $2.12_{-0.06}^{+0.02}$ & $\sim0$ & 1.07(2671)
\\
901 & $ 3.11_{-0.03}^{+0.04} $ & $ 1.27_{-0.01}^{+0.01} $ & $ ^{\mathrm{a}}1.90_{-0.02}^{+0.08} $ & $ 2.60_{-0.01}^{+0.01} $ & $ 94.62_{-2.30}^{+3.55} $ & $ 83.72_{-2.22}^{+2.69} $ & $14.65_{-0.38}^{+0.47}$ & $0.48_{-0.11}^{+0.15}$ & $3.76_{-0.39}^{+0.21}$ & $\sim0$ & 1.04(2827)
\\
902 & $ 3.44_{-0.13}^{+0.01} $ & $ 1.26_{-0.03}^{+0.01} $ & $ ^{\mathrm{a}}1.77_{-0.07}^{+0.04} $ & $ 2.76_{-0.02}^{+0.01} $ & $ <147 $ & $ 117.69_{-1.18}^{+2.59} $ & $15.16_{-0.39}^{+0.45}$ & $0.42_{-0.03}^{+0.02}$ & $5.85_{-0.33}^{+0.02}$ & $\sim0$ & 1.20(2827)
\\
903 & $ 3.04_{-0.02}^{+0.01} $ & $ 1.29_{-0.00}^{+0.01} $ & $ ^{\mathrm{a}}1.63_{-0.08}^{+0.01} $ & $ 2.55_{-0.02}^{+0.00} $ & $ 82.20_{-2.73}^{+4.24} $ & $ 79.85_{-2.05}^{+0.57} $ & $14.23_{-0.62}^{+0.07}$ & $1.28_{-0.01}^{+0.06}$ & $3.41_{-0.10}^{+0.19}$ & $\sim0$ & 0.94(2743)
\\
\hline
\end{tabular}
\end{center}
\begin{list}{}{}
\item[${\mathrm{a}}$]{: $(10^{3})$}
\item[${\mathrm{b}}$]{: The normalization of $diskbb$ model.}
\end{list}
\end{table*}

Figure 5 shows that along the spectral evolution towards SIMS, the FRS residual tends to appear gradually at energies above a few tens of keV, which may require an additional spectral component that has less contribution to QPO FRS. We hence add a $powerlaw$ component to the previous fitting, and thus have model-3 written as:

\begin{equation}
\rm rms = \sigma_{1}(\emph E) \times \emph f_{1}^{\prime} + \sigma_{2}( \emph E) \times\ \emph f_{2}^{\prime},
\end{equation}
\begin{equation}
f_{1}^{\prime} = \frac{\emph F_{\rm pl}(\Gamma_{1},\emph E_{\rm cut},n_{1})}{\emph F_{\rm d}(\emph T,n_{2})+\emph F_{\rm pl}(\Gamma_{1},\emph E_{\rm cut},n_{1})+\emph F_{\rm pl}^{\prime}(\Gamma_{2},n_{3})},
\end{equation}
\begin{equation}
f_{2}^{\prime} = \frac{\emph F_{\rm d}(\Gamma_{1}, n_{2})}{\emph F_{\rm d}(\emph T, n_{2})+\emph F_{\rm pl}(\Gamma_{1}, \emph E_{\rm cut}, n_{1})+\emph F_{\rm pl}^{\prime}(\Gamma_{2}, n_{3})},
\end{equation}
\begin{equation}
\sigma_{1}(\emph E) = \frac{\sigma_{0}}{1+e^{-k(E-E_{0})}},
\end{equation}
\begin{equation}
\sigma_{2}(\emph E) = \sigma_{\rm disk},
\end{equation}
where $F_{\rm pl}^{\prime}$ is the flux of the new power-law component which does not contribute to the rms, and $\Gamma_{1}$ is the photon index of cutoff power-law component. $\Gamma_{2}$ is the photon index, and $n_{3}$ is the normalization of the new power-law component added in model-3.
For convergency of the joint fitting, we fix the values $\sigma_{0}$, $k$, $E_{0}$ and $\sigma_{\rm disk}$ for those in Table 5 and obtain new outputs listed in Table 6.
Correspondingly, the spectral residuals are shown in Figure 5 in red triangle points.
With model-3 the FRS residual at high energies alleviates and the reduced $\chi^{2}$ decreases compared to the previous joint fitting. As shown in Figure 4, with model-3, the overall evolution of the spectral parameters trend changed a little and the disk normalization derived in ObsIDs 145-501 drops from an average of 1141.63 to 696.41, and the latter value is more consistent with the results in \cite{Tao et al. 2018} during this time period. Also the corona temperature is similar to those reported with NuSTAR observations during the LHS.
\begin{table*}[ptbptbptb]
\begin{center}
\label{table}
\caption{The joint fit with disk component by model-3:}
\begin{tabular}{c|cccccccc|c}
\hline
\hline
ObsID & $N_{\rm H}$ &  $kT_{\rm in} $ & $^{b}n_{2}$ & $\Gamma_{1}$ & $E_{\rm cut} $ & $^{b}n_{1}$ & $\Gamma_{2}$ & $^{b}n_{3}$ & $\chi_{\rm red}^{2}(d.o.f)$
\\
& $(10^{22}\rm cm^{-2})$ & (keV) & $(10^{5})$ & & (keV) & & &$(10^{-4})$ &
\\
\hline
144 & $ 3.54_{-0.20}^{+0.18} $ & $ 0.37_{-0.01}^{+0.01} $ & $ 2.80_{-0.45}^{+0.77} $ & $ 2.19_{-0.01}^{+0.01} $ & $ 50.95_{-1.03}^{+1.14} $ & $ 37.17_{-0.66}^{+0.73} $ & $ 0.19_{-0.10}^{+0.10} $ & $ 1.14_{-0.50}^{+0.71} $ & 1.09(2830)
\\
145 & $ 4.07_{-0.19}^{+0.05} $ & $ 0.35_{-0.01}^{+0.01} $ & $ 6.78_{-0.49}^{+0.16} $ & $ 2.24_{-0.01}^{+0.01} $ & $ 51.76_{-0.83}^{+0.68} $ & $ 42.04_{-0.74}^{+0.22} $ & $ 0.83_{-0.03}^{+0.04} $ & $ 31.33_{-5.65}^{+7.71} $ & 1.19(2830)
\\
301 & $ 3.72_{-0.27}^{+0.28} $ & $ 0.36_{-0.01}^{+0.01} $ & $ 4.31_{-1.09}^{+1.60} $ & $ 2.04_{-0.02}^{+0.01} $ & $ 37.42_{-1.38}^{+0.32} $ & $ 34.53_{-1.16}^{+1.02} $ & $ 1.00_{-0.05}^{+0.06} $ & $ 134.66_{-32.20}^{+48.62} $ & 0.97(2822)
\\
401 & $ 4.04_{-0.25}^{+0.13} $ & $ 0.35_{-0.01}^{+0.01} $ & $ 7.42_{-0.80}^{+0.78} $ & $ 2.26_{-0.01}^{+0.01} $ & $ 46.69_{-1.38}^{+0.66} $ & $ 54.04_{-1.26}^{+0.45} $ & $ 1.10_{-0.05}^{+0.04 }$ & $ 133.85_{-27.20}^{+33.40} $ & 1.03(2830)
\\
501 & $ 3.86_{-0.20}^{+0.12} $ & $ 0.33_{-0.01}^{+0.01} $ & $ 7.96_{-1.70}^{+1.21} $ & $ 2.36_{-0.01}^{+0.01} $ & $ 51.51_{-1.21}^{+0.72} $ & $ 72.95_{-1.91}^{+0.72} $ & $ 0.67_{-0.06}^{+0.04} $ & $ 17.43_{-4.71}^{+3.31} $ & 1.13(2830)
\\
601 & $ 3.75_{-0.30}^{+0.24} $ & $ 0.42_{-0.02}^{+0.02} $ & $ 1.78_{-1.06}^{+0.40} $ & $ 2.26_{-0.02}^{+0.01} $ & $ 43.27_{-1.95}^{+1.05} $ & $ 62.03_{-2.17}^{+0.62} $ & $ 0.99_{-0.06}^{+0.09} $ & $ 88.59_{-21.55}^{+47.29} $ & 1.01(2673)
\\
901 & $ 2.92_{-0.05}^{+0.01} $ & $ 1.27_{-0.01}^{+0.01} $ & $ ^{\mathrm{a}}2.03_{-0.05}^{+0.08} $ & $ 2.48_{-0.02}^{+0.01} $ & $ 57.07_{-2.86}^{+0.84} $ & $ 66.38_{-3.85}^{+0.28} $ & $ 1.23_{-0.05}^{+0.08} $ & $ 181.99_{-31.25}^{+63.40} $ & 0.99(2829)
\\
902 & $ 2.97_{-0.02}^{+0.04} $ & $ 1.27_{-0.01}^{+0.01} $ & $ ^{\mathrm{a}}2.03_{-0.03}^{+0.09} $ & $ 2.52_{-0.01}^{+0.01} $ & $ 61.02_{-1.67}^{+1.37} $ & $ 71.47_{-2.19}^{+0.33} $ & $ 0.73_{-0.04}^{+0.06} $ & $ 23.75_{-4.62}^{+6.30} $ & 0.99(2829)
\\
903 & $ 2.92_{-0.04}^{+0.03} $ & $ 1.28_{-0.01}^{+0.01} $ & $ ^{\mathrm{a}}1.78_{-0.07}^{+0.05} $ & $ 2.45_{-0.01}^{+0.01} $ & $ 49.91_{-2.11}^{+2.38} $ & $ 68.78_{-1.66}^{+1.42} $ & $ 1.01_{-0.05}^{+0.09} $ & $ 95.21_{-16.81}^{+57.80} $ & 0.89(2745)
\\
\hline
\end{tabular}
\end{center}
\begin{list}{}{}
\item[${\mathrm{a}}$]{: $(10^{3})$}
\item[${\mathrm{b}}$]{: The normalization of $diskbb$ model.}
\end{list}
\end{table*}

\begin{figure*}[htbp]
\centering\includegraphics[scale=1]{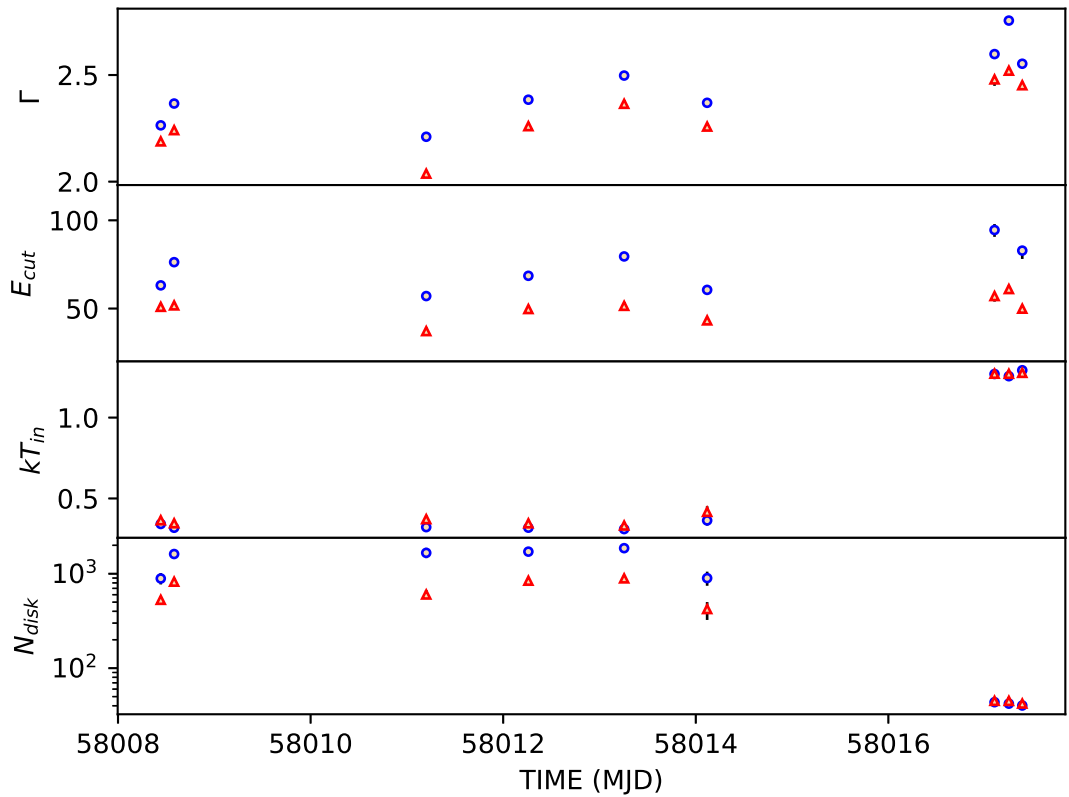}
\caption{The blue points and red points are the parameters of energy spectral fittings and joint fitting respectively. The blue points are taken from Table 1, and the red points are taken from Table 6.
}
\end{figure*}

\begin{figure*}[htbp]
\centering\includegraphics[scale=0.7]{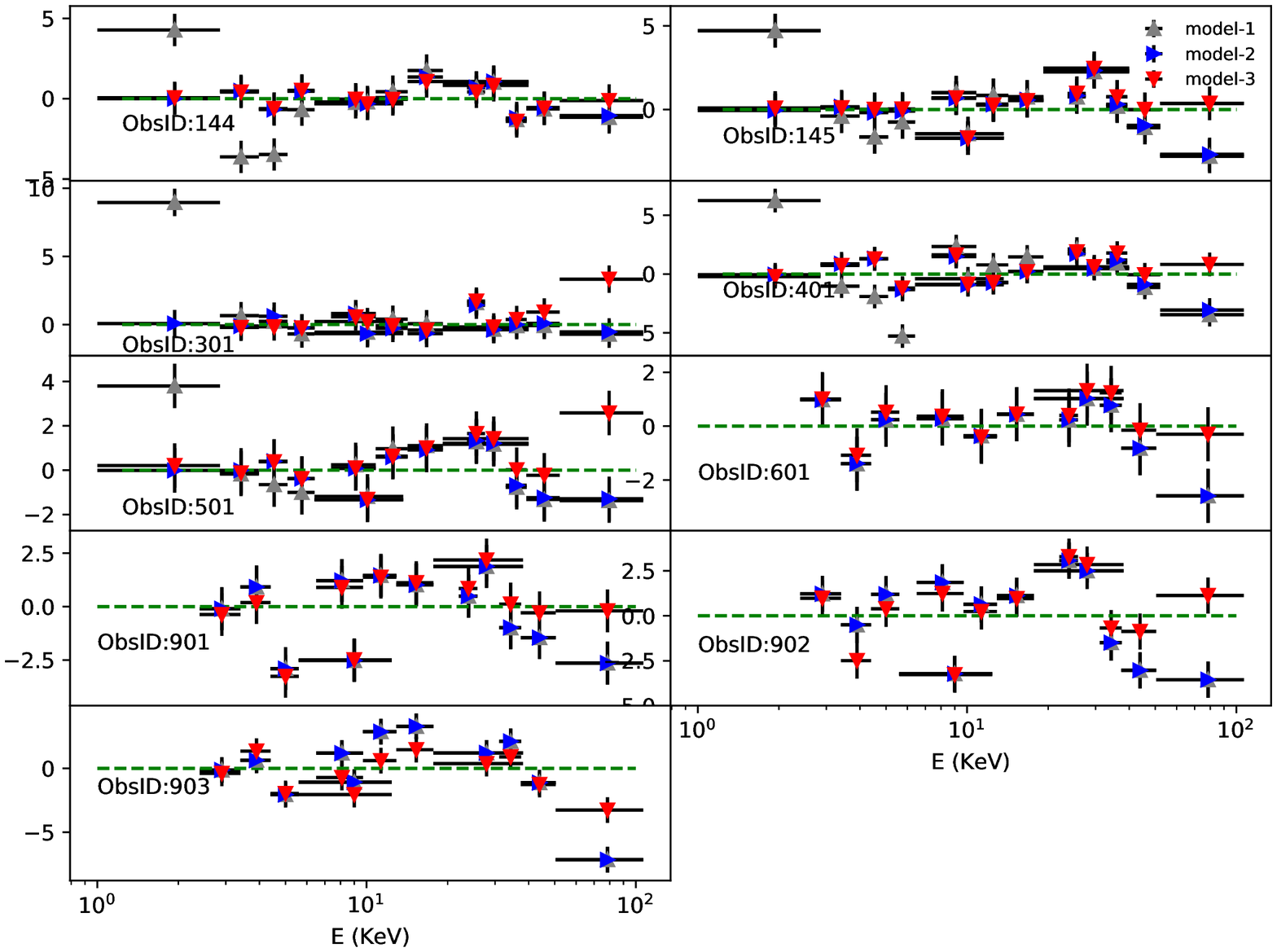}
\caption{The panels show the residuals ((data-model)/error) of QPO FRS fittings with model-1 (grey triangle), model-2 (blue triangle) and model-3 (red triangle). Compared with ObsID: 601, 901, 902 and 903, the other ObsIDs reveal larger differences in low energy range while model-2 gives better residuals than model-1. Compared with model-1 and model-2, model-3 gives a fit in the high energy range.}
\end{figure*}

We plot the intrinsic QPO FRS of the non-thermal component in Figures 6c and 6d, derived from the joint fitting with model-1 and model-2.
\begin{figure*}[htbp]
\centering\includegraphics[scale=0.6]{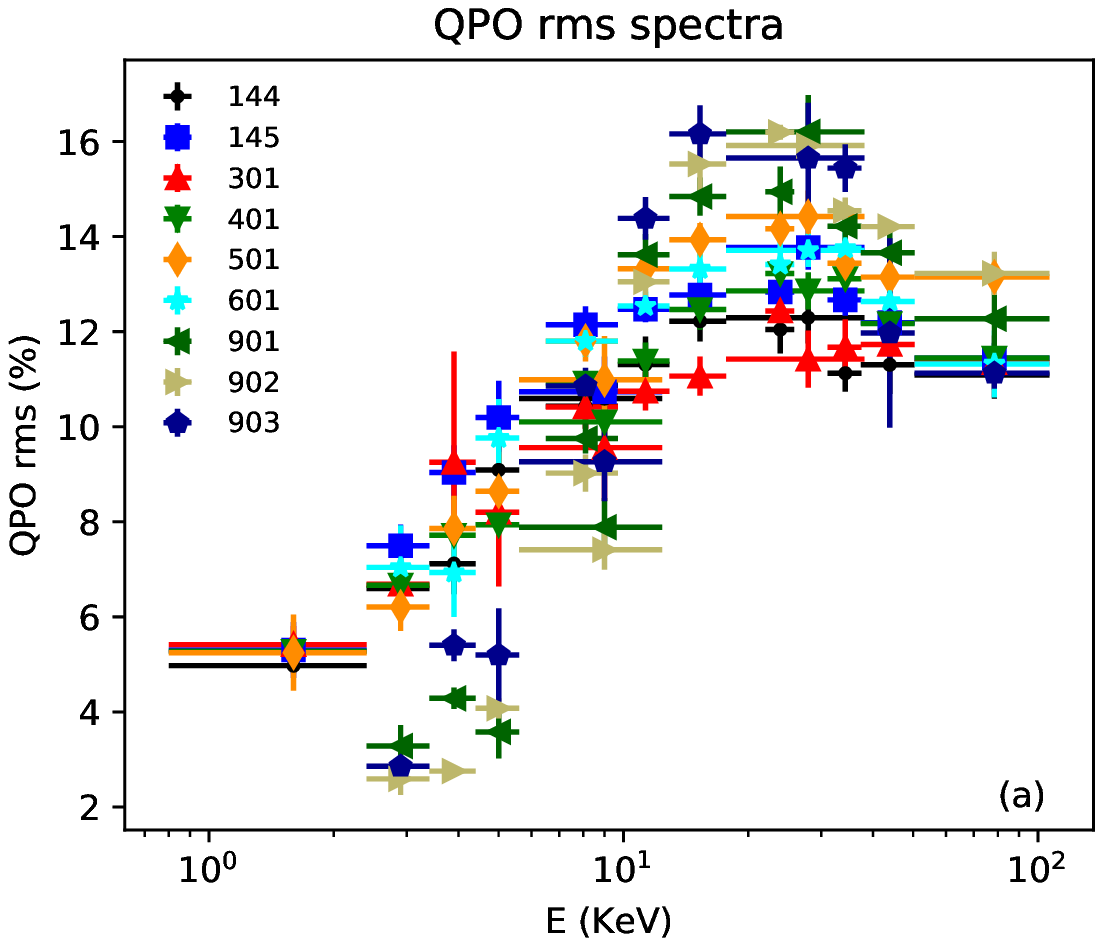}
\centering\includegraphics[scale=0.5]{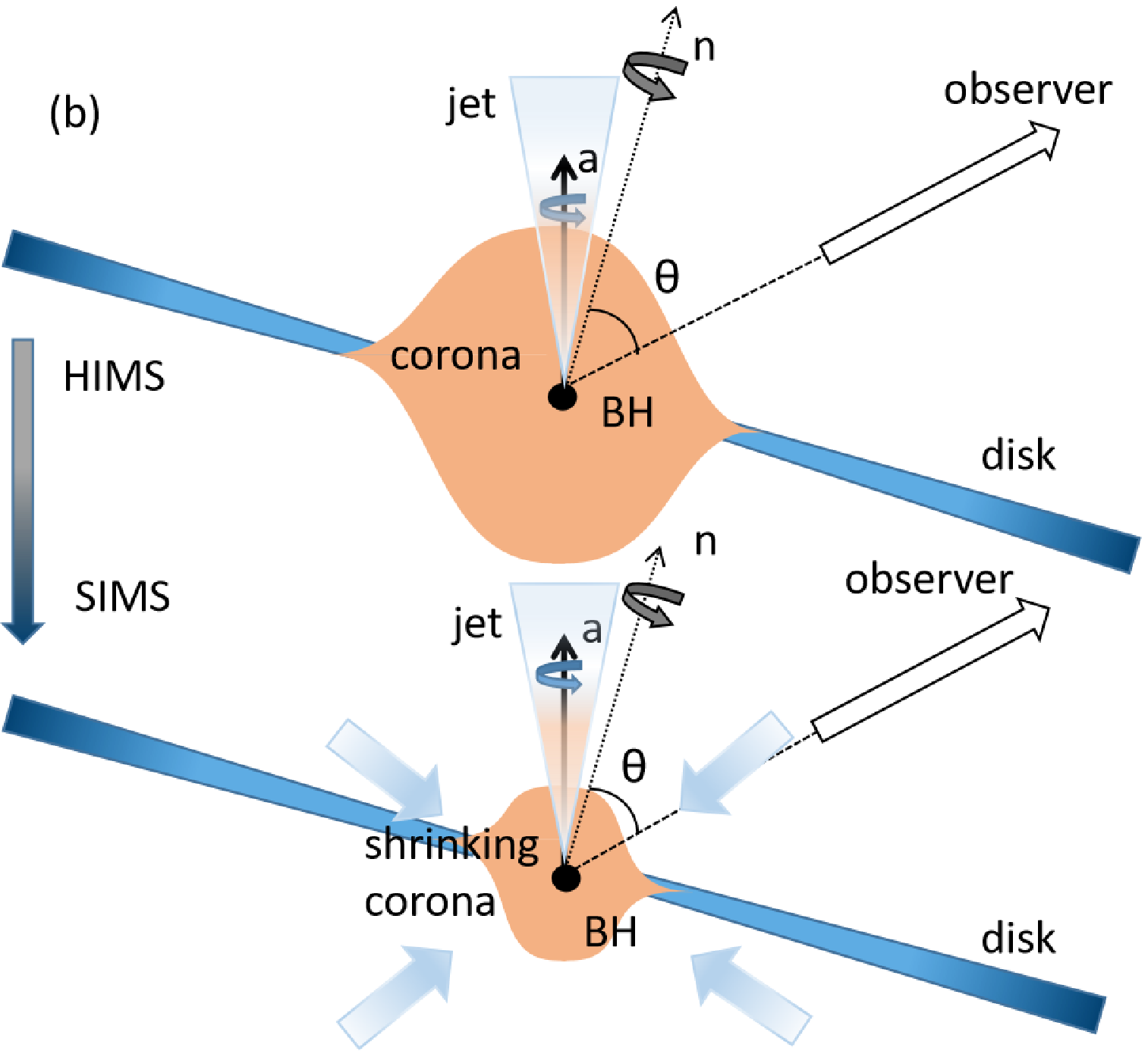}
\centering\includegraphics[scale=0.7]{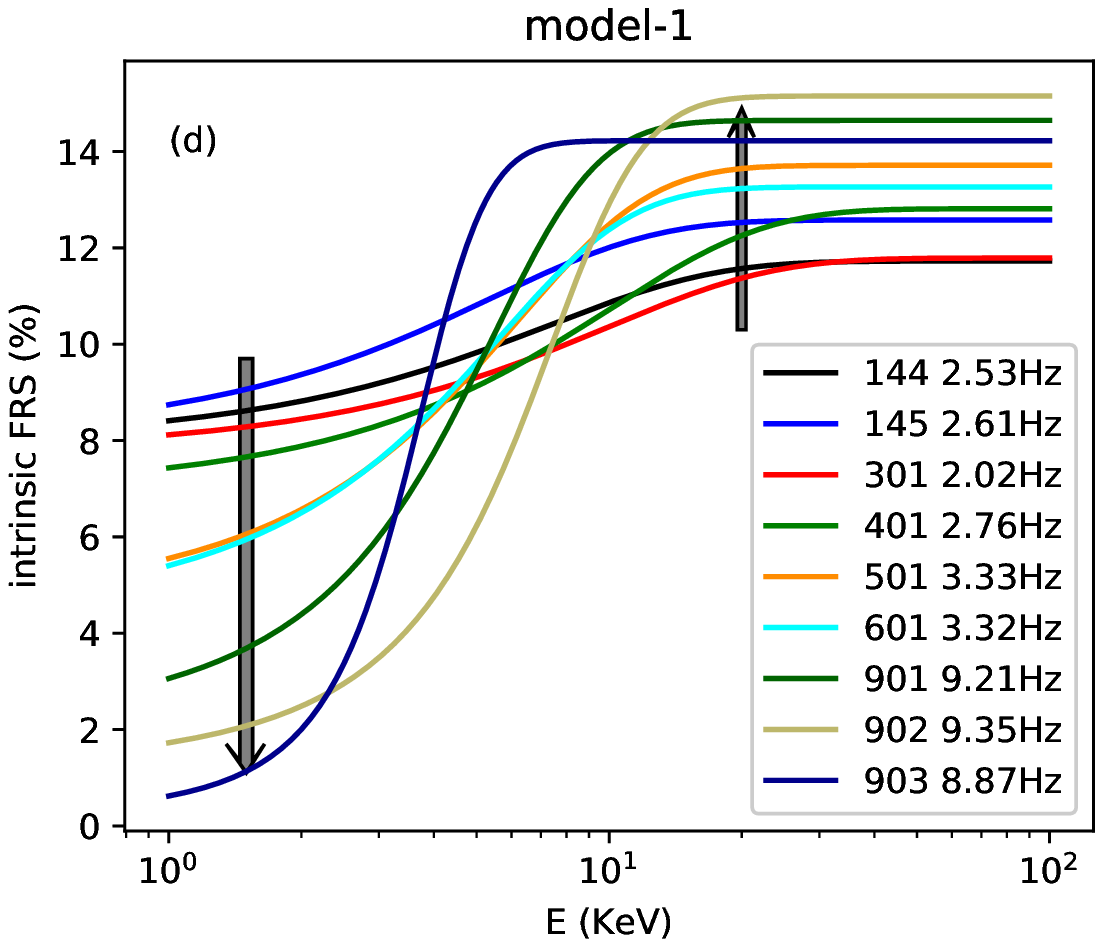}
\centering\includegraphics[scale=0.7]{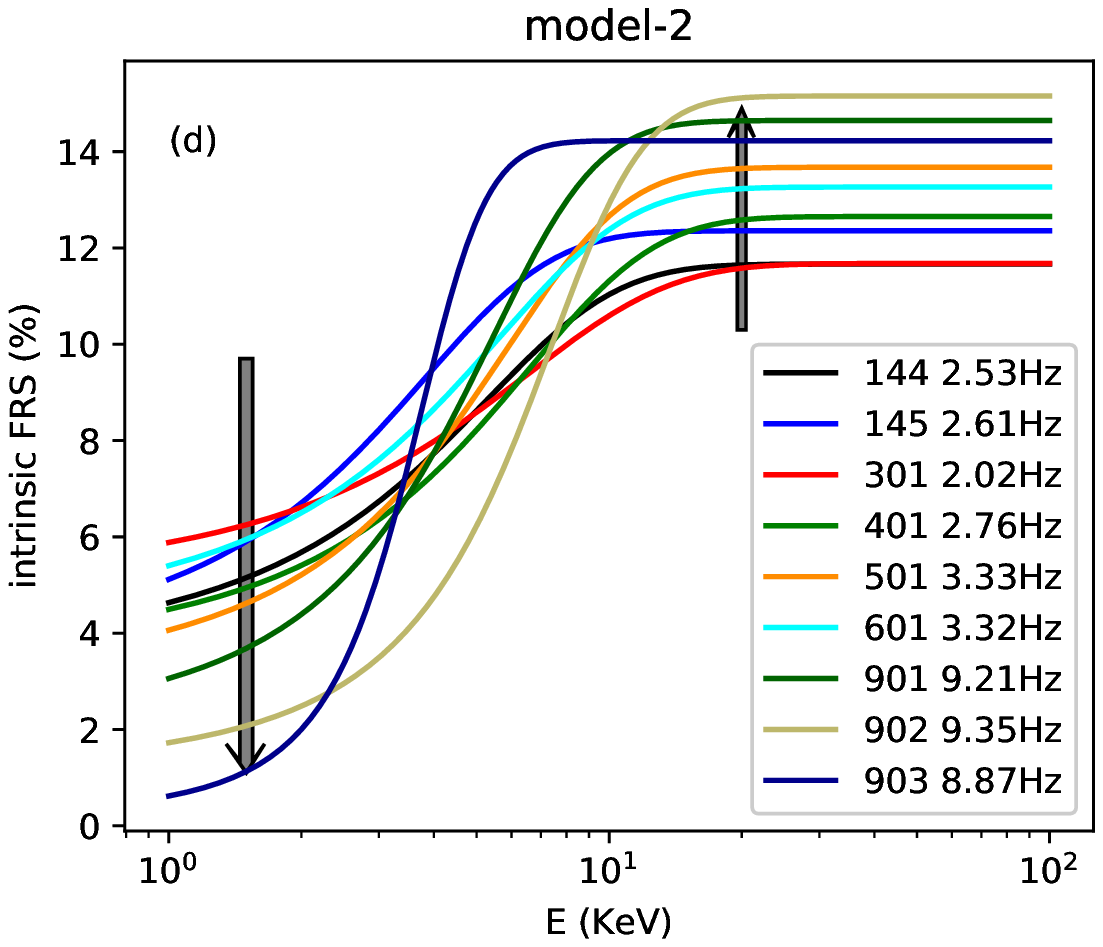}
\caption{(a): QPO RFS of different ObsIDs. Form ObsID: 144 to ObsID: 903, the states evolve from HIMS to SIMS.
(b): Illustrattion of the corona evolving along with the outburst during HIMS and SIMS. \textbf{a} is the spin of black hole, \textbf{n} is the angular momentum of the accretion flow, and $\theta$ is the inclination angle.
The intrinsic FRS $\sigma(E)$ derived from joint fittings with model-1 (panel c) and model-2 (panel d). The grey arrow displays the evolutionary trend from HIMS to SIMS.}
\end{figure*}
Although the QPO FRS in HIMS and SIMS have similar overall evolution trend at the first glance, the joint fittings reveal a difference between them. From HIMS to SIMS, the QPO FRS decreases at soft X-rays and increases at hard X-rays, which suggests a hardening of the QPO FRS.
\section{Discussion}
We performed a spectral and timing analyses for the outburst of MAXI J1535-571 in a broad energy band as observed by \emph{Insight}-HXMT during the initial low hard state and the intermediate state. Although results from the energy spectral fittings are in general consistent with what was previously reported by Swift/XRT at soft X-rays and by NuSTAR in the LHS, we find that in the intermediate state the energy spectrum becomes softer, and the cutoff energy extends towards  the higher energies. The intrinsic QPO FRS is defined as the variability amplitude of the source flux, and spectral components that cannot be disentangled in the energy domain may be distinguishable in the time domain due to the difference in the QPO rms contributions. The joint energy spectrum and QPO fractional rms spectrum (FRS) fitting is, according to the definition of FRS, to take the different spectral components as inputs to probe the intrinsic QPO FRS. Here the energy spectral components are classified into thermal and non-thermal, where the non-thermal component can either be presented by phenomenological model like cutoff power-law as adopted in this paper or the more physical models. Apart from having the intrinsic QPO FRS shape as the main focus of this paper, analyzing further the FRS evolution with parameters of the energy spectrum would need more physical spectral model in the joint fitting. We note that for the observations where the joint spectral fittings are performed with type-C QPO FRS, the reflection component is not prominent, probably due to that the \emph{Insight}-HXMT data in intermediate hard state are relatively poorer for LE at above 6 keV and for ME at above 20 keV, which are critical for constraining the reflection components. The QPO FRS is found to be dominated by the non-thermal component and becomes harder as the outburst evolves (Tabel 5).

The joint spectral analysis was carried out in different ways previously (\citealp{GZ2005}; \citealp{You2018}; \citealp{SZ2006}). The FRS integrated over (1/512)-128 Hz of the total power spectrum (i.e. without QPO component) was investigated by \cite{GZ2005}.
They analyzed the RXTE data and fitted the energy spectra with a Comptonization plus a disk blackbody model representing non-thermal emission plus thermal emission from the disk.

They took these spectral components as inputs to estimate the FRS and assigned each component a constant to account for the corresponding contribution to the overall flux variability. They found that the FRS can be recovered if the variability of the spectral parameters in each component was taken into account. As a result, in the hard state, the flux variability as seen in the continuum power spectrum is mostly contributed by variation of the disk emission, and the fluctuation of the non-thermal emission might take part in it. Our results show that the intrinsic variability fraction of the energy spectral components are energy dependent and cannot be solely represented by constants. Since the FRS of the total power spectrum is usually very different from that of the QPOs, the QPOs may have an origin different from the continuum power spectrum. We therefore adopted a joint fitting procedure different from the above method to the type-C QPOs, which gives more proper inputs from the energy spectral components.

An alternative to understanding the origin of the type-C QPOs is the
L-T precession scenario (\citealp{Ingram et al. 2009}). The
accreting material orbiting in the vicinity of the black hole will
precess due to the mis-alignment of the angular momentum between the
spinning black hole and the accretion disk. The periodic
change in projection of the emission region to the line of sight
of the observer will give QPO features in power spectra. Such
process was studied in detail by \cite{You2018} via light-tracing
simulations. They considered a system with a truncated disk and a
precessing corona in the hard state. The soft X-rays produced in the
disk were traced all the way throughout their journey to the corona,
where the Compton scattering, relativistic light bending effect and
the effect of inner wrapped disk were properly accounted for. Their
results show that the L-T precession can lead to an FRS shape similar
to what we measure in MAXI J1535-571 (i.e. the FRS increases
gradually with energies and becomes flat at hard X-rays).
However, they did not perform a joint fitting with the energy
spectra.

Multiplication of the FRS with the total flux can result in an energy spectrum for a specific given frequency range.  In such a way \cite{SZ2006} obtained the energy spectra for QPOs detected in 0.1 - 10 Hz and performed spectral fittings using XSPEC for a few BHXRBs. They found that the energy spectra extracted around the QPO frequencies are anti-correlated with the time-averaged ones, and the disk contribution to the QPO rms is absent.
In addition, the ratio of the energy spectra around the QPO frequencies to the non-thermal spectral component is energy dependent.
Such a ratio correspondence to the rms function of $\sigma(E)$ is introduced in models 1-3 (this study), where the contribution from the disk emission to the fractional rms is fully considered.

The joint spectral fitting on MAXI J1535-571 also gives similar results: during the intermediate state, the intrinsic fractional variability of the type-C QPO becomes harder and the disk contribution to the rms disappears when the energy spectrum gets softer.
We note that one type-B QPO was reported in observation between Obs. 601 and 901 (\citealp{Huang et al. 2018}), therefore, the source may tend to evolve towards SIMS afterward.
Moreover, we find that as the outburst evolves towards the soft intermediate state, the fractional variability of the QPOs increases at hard X-rays and decreases at soft X-rays.
We speculate that, in an L-T precession scenario, such an evolution of the QPO FRS may be related to a corona cooling process by the disk thermal emission. As a result, a cooler corona may have a relatively smaller size, with a smaller outer radius and a flatter inner shape since it is generally believed that the temperature goes up in the inner part of the corona. As shown in Figures 6c and 6d, the intrinsic rms of the non-thermal emission component increases and reaches a flat top at an energy around 10 keV.
This may be understood if the outer part of the corona has a geometric shape different from the inner part: as illustrated in Figure 6b, since the corona originally comes from the accretion matter of the inner disk, there may exist a region for the outer part of the corona to connect to the inner part of the disk.

Once the corona is cooler, it shrinks to a smaller size. Such
evidence was reported in MAIXJ1820+070 by \cite{Kara2019}, where the
corona height with respect to the black hole was observed to
decrease along with the outburst evolution. The disk-corona
connection part which corresponds to the QPO FRS at soft X-rays may
become steeper with respect to the line of sight and hence results
in less projection effect in L-T precession.

The FRS residuals as shown at higher energies in the joint fitting may indicate two possibilities: either the energy spectrum has an additional component or the intrinsic QPO FRS has to turn over at high energies along with evolution of the outburst.
By introducing an power law as an additional spectral component to account for the QPO FRS residuals at high energies, we find that, the spectral parameters of this additional varies a lot. As sown in Table 6, the power law normalization can vary by factor of 100, which is not likely realistic. Also if assigns such an additional power law component to jet, it would then be not consistent with the fact that a jet contribution usually becomes small when the energy spectrum softens.
An alternative consideration to account for the QPO FRS residuals is that, the `sigmoid' function is not sufficient to recover the QPO FRS and hence the observed FRS residuals in the joint fitting are intrinsic to the QPO FRS. The intrinsic QPO FRS has the trend to turn over at high energies along with evolution of the outburst. In a L-T precession scenario, this may indicate that the precessing  inner hot flow may have a complex shape, where the L-T procession motion could become weaker with a smaller misaligned angle between the precessing inner material and the spin of the BH (\citealp{HM2006}).
\section{Summary}
The broad energy coverage and large effective area of
\emph{Insight}-HXMT allow us to investigate the outburst behaviors in time and energy domains jointly.
This joint fitting approach on the newly discovered black hole candidate MAXI J1536-571 reveals a few interesting results.
During the intermediate state, the type-C QPOs show a peculiar evolution in the intrinsic rms spectrum, which may be related to the shrinking of the corona cooled by the disk thermal emission.
The residuals showing up in the joint fitting may indicate either an additional spectral component or  a turn-over trend intrinsic to QPO FRS at high high energies.

\acknowledgements
This work is supported by the National Key R\&D Program of China (2016YFA0400800) and the National Natural Science Foundation of China under grants 11733009, U1838201 and U1838202. This work made use of data from the \emph{Insight}-HXMT mission, a project funded by China National Space Administration (CNSA) and the Chinese Academy of Sciences (CAS).

\bibliographystyle{plain}

\end{document}